\documentstyle[bo99,epsfig]{article}

\title{Constraints to the SSC model for Mkn 501}
\author{Fabrizio Tavecchio and Laura Maraschi}
\affil{ Osservatorio Astronomico di Brera, Milano, Italy}

\begin{document}

\maketitle

\begin{abstract}
We fit the SEDs of the TeV blazar Mkn 501 adopting the homogeneous 
Synchrotron-Self Compton model to simultaneous X-ray and TeV spectra recently 
become available.
We present detailed model spectra calculated with the above constraints and 
taking into account the absorption of TeV photons by the IR background. We 
found that the curved TeV spectra can be naturally reproduced even without 
IRB absorption. Taking IRB absorption into account changes the required parameter
values only slightly.
\end{abstract}

\section{Introduction}

The study of Blazars has been recently enriched by the detection of very
high energy gamma rays (energy in the TeV range) from a handful of nearby
sources. This discovery opens the possibility to effectively test and 
constrain the radiative mechanisms invoked to explain the emission from 
Blazars and to investigate the physical conditions in relativistic jets.

As discussed in Tavecchio et al. 1998 (hereafter T98), the knowledge of the simultaneous X-ray 
and TeV spectra of Blazars allows to univocally find the set of 
physical parameters necessary within  a homogeneous Synchrotron--Self Compton 
model with an electron distribution described by a broken power-law.
Here we apply the SSC model to recent high quality, simultaneous  
X-ray and TeV data of the well studied source Mkn 501 and discuss the results. 

\section{The data}

\noindent
TeV spectra measured by the CAT team during 1997 were recently reported 
in Djannati-Atai 
et al (1999). In particular the TeV spectra associated with 
$Beppo$SAX observations of April 16 and 7 (reported in Pian et al 1998) are also
discussed. For the large flare of April 16, given the high flux level, 
it has been possible to obtain a good
quality TeV spectrum from a single observation which partially overlaps with the $Beppo$SAX
observation. On Apr 7, Mkn 501 was less bright and the TeV spectrum is 
obtained using different observations
with similar TeV fluxes and hardness ratios (for more details see Djannati-Atai 
et al 1999).
We combine these data with the X-ray spectra reported 
in Pian et al. (1998) for the same days constructing quasi simultaneous SEDs for two
epoches. The TeV spectra of Mkn 501 are clearly curved. It was suggested that 
this curvature could be the "footprint" of the absorption of TeV photons by
the Infra Red Background (IRB, e.g Konopelko et al. 1999, but see
the detailed discussion in Vassiliev 1999)

\section{Spectral Fits with the SSC model}

T98 obtained analytical relations for
the IC peak frequency in Klein-Nishina (KN) regime adopting the step function approximation for 
the KN cross-section. A comparison between these approximate estimates and the 
detailed numerical calculations with the full KN cross section shows that the analytical formulae
for the IC peak frequency given in T98 can overestimate its value 
by a factor of 3-10, depending on $n_2$, the index of the steeper part of the electron distribution. Therefore,
although the analytical estimates are useful as a guideline, 
it is very important for the study of the TeV spectrum of Blazars to use
precise numerical calculations, like those discussed in the following.\\

\begin{table}
\begin{center}
\begin{tabular}{ccccccc}
\hline
\hline
$R_{16}$ & B& $\delta$ & $\gamma _{break}$ & $K$ &$n_1$  &  $n_2$\\
(cm)& (G)& & & & & \\ \hline
\multicolumn{7}{c}{ {Mkn 501 High state (Apr 16)}} \\
\hline
1.8 & 0.02 & 10 & $10^7 $ & $10^4$ & 2 & 8\\
\multicolumn{7}{c}{{Mkn 501 High state with IRB}} \\
\hline
3.5 & 0.005 & 10 & $2\times 10^7 $ & $4\times 10^3$ & 2 & 8\\
\multicolumn{7}{c}{ {Mkn 501 Low state (Apr 7)}} \\
\hline
2.5 & 0.01 & 10 & $6.3\times 10^6 $ & $2.3\times 10^3$ & 2 & 8\\
\hline
\end{tabular}
\caption{Values of the physical parameters used for the SSC model.}
\end{center}
\end{table}

We reproduced the SED of Mkn 501 (reported in Fig.1) using the SSC model described 
in T98 and Chiappetti et al 1999. In a spherical region with radius R, magnetic field B and Doppler
factor $\delta $, a population of electrons with energy distribution of
the form $N(\gamma )=K\gamma ^{-n_1}(1+\gamma/\gamma _{break})^{n_1-n_2}$
emits synchrotron and IC radiation. 
The IC spectrum is calculated with the full Klein-Nishina cross-section, using 
the formula derived by Jones (1968). The seed photons for the IC scattering 
are those produced by the same electrons through the synchrotron mechanism.
For the case of the high state of Mkn 501 we considered possible intervening
absorption by the IRB adopting the prescriptions for the {\it low model} 
discussed in Stecker \& De Jager (1998). 
In the calculation reported here we assume a typical radius of the emitting
region of $R\sim 10^{16}$ cm and a Doppler factor $\delta =10$. With this choice 
the minimum variability timescale is of $\simeq 10$ h (see Tab.1 for the parameters
used).

It is useful to calculate the IC emission from electrons with different
energy ranges. These "slices" are reported in Fig. 2 (see figure caption for
the detailed energy ranges). It is interesting to note
that, because of the KN cut at the high energy emission, the peak of the IC
component is produced by electrons with Lorentz factor below $\gamma _{break}$
while the peak of the Synchrotron component is produced by electrons at 
$\gamma _{break}$. This effect has important consequences for the 
study of the X-ray/TeV correlated variability (see also Maraschi et al. 1999).

\begin{figure}
\vskip -3 cm 
\centerline{\psfig{figure=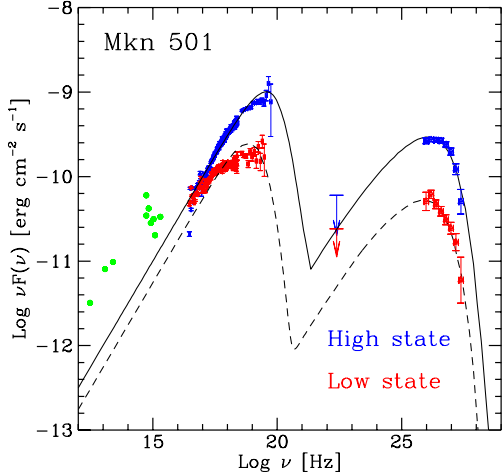,width=8 truecm,clip=}}
\vspace*{-0.8cm}
\caption{SED of Mkn 501 (High state and
Low state) with the spectrum calculated with the SSC model. Upper limits at 100 
MeV band are from Weekes et al. 1996 and Catanese et al 1997.}
\end{figure}

\section{Discussion and Conclusions}

The parameters adopted for the model are reported in Table 1. 
Our model indicates a rather low value for the magnetic field and a high 
$\gamma _{break}$ (see Tab.1). The transition from the low state to the high state in Mkn 501 is 
consistent with an increase of a factor of 2 in both $\gamma _{break}$ and B and with an almost constant
value of $\delta $ and $R$. These results are in agreement with a similar 
analysis of the April 16 flare by Bednarek \& Protheroe (1999): for a minimum timescale of 2.5 h they
found that the TeV spectrum can be well reproduced by $B\sim 0.03$ and $\delta
\sim 15$.

Although some authors (see Konopelko et al. 1999) have recently proposed
that the curvature of the TeV spectrum of Mkn 501 provides evidence for absorption by IRB, the curvature
could be intrinsic and related to the curved electron distribution necessary
to fit the X-ray data. Our models
show that a curved spectrum is a plausible explanation for the curvature.
The introduction of the IRB does not dramatically change the inferred
physical parameters. As shown in Fig.2 IRB/no IRB TeV spectra are very similar
up to 15 TeV, while for higher energy the power-law spectrum of the no IRB case
is changed in an exponential profile. Therefore spectral data above 10 TeV 
are needed in order to understand if IRB absorption affects Blazar spectra.

Finally we note that in the theory
of particle acceleration by shocks (for a recent critical discussion see Henri 
et al. 1999) the maximum Lorentz factor of accelerated electrons, obtained
equating acceleration time and cooling time, is given by $\gamma _{max}\simeq 
10^7 (B/0.02 G)^{-1/2}$, where we used the value of B we found in the high state
of Mkn 501. Therefore our fit suggests that during the high activity states
of Mkn 501 the
electrons can reach the maximum energy fixed by the balance between cooling
and acceleration processes. In states of lower activity the acceleration time
could be longer and the condition $t_{esc}<t_{acc}$ (where $t_{esc}$ and 
$t_{acc}$ are the escape time and the acceleration time respectively) could 
prevent the electrons for reaching the maximum energy.

\begin{figure}
\vskip -4 cm
\hskip 0.5 truecm
\centerline{\psfig{figure=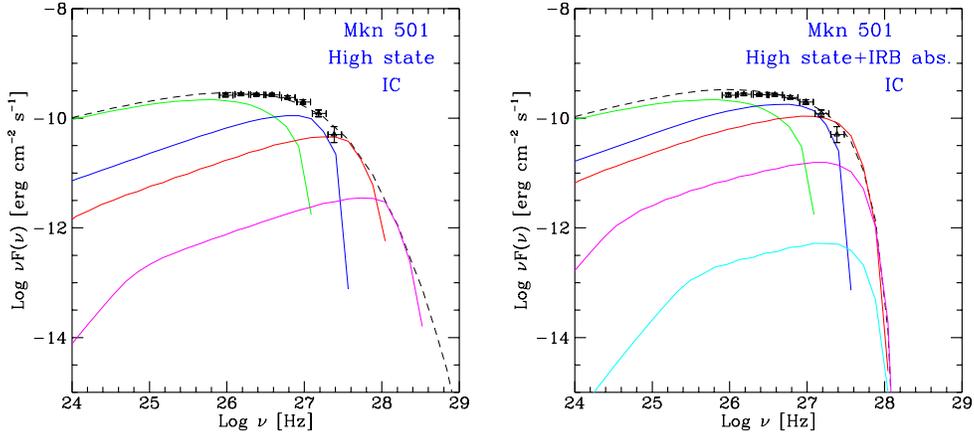,width=11 truecm,clip=,angle=-90}}
\vspace*{-3.0cm}
\caption{TeV spectrum of Mkn 501-High state without (left) and with (right) 
IRB absorption. We also show the contribution from electrons with different Lorentz 
factors. From the left the curves show the emission from electrons with Lorentz factor
in the range: $1-10^6$, $10^6-3\times10^6$, 
$3\times10^6-10^7$, $10^7-3\times10^7$, $3\times10^7-10^8$}
\end{figure}

\begin{acknowledgements}
We thank A. Djannati-Atai for sending us the TeV data of 
Mkn 501.
\end{acknowledgements}

\end{document}